\def\comment#1{}
\def\lfrac#1#2{#1/#2}
\def\betan{\mu}

\def\cm#1{}

 \def\lfrac#1#2{{{{#1}/{#2}}}}

\documentstyle[pra,epsf,aps]{revtex}
\begin{document}
\title{
Path integral for a relativistic Aharonov-Bohm-Coulomb system
}
\author{
 De-Hone Lin%
 \thanks{e-mail:d793314@phys.nthu.edu.tw}}
\address{
 Department of Physics, National Tsing Hua University,\\
Hsinchu 30043, Taiwan, Republic of China%
}
%%%%%%%%%%%%
\maketitle
\begin{abstract}
The path integral for the relativistic spinless Aharonov-Bohm-Coulomb system
is solved, and the energy spectra are extracted from the resulting amplitude.
\end{abstract}

\section{INTRODUCTION}

With the help of Duru and Kleinert's
 path-dependent time transformation
\cite{DK} the list of solvable path integrals
has been extended to
essentially all
potential problems
which possess a solvable
Schr\"odinger equation
 \cite{1,2}.
Only recently has the technique been
extended to relativistic
potential problems \cite{7}, followed by two
applications \cite{5,6,8.1,8}.
Here we'd like to add a further application
by solving the path integral
of relativistic particle
in two dimensions in the presence of an
infinitely thin
Aharonov-Bohm
magnetic field along the $z$-axis  \cite{9}
and a $1/r$-Coulomb potential (ABC system).
This may be relevant for understanding the behavior of
relativistic charged anyons which are restricted to a plane
but whose Coulomb field extends into
three dimensions\cite{1,15}.

\section{THE RELATIVISTIC PATH INTEGRAL
}
Adding
a vector potential ${\bf A}({\bf x})$
to Kleinert's relativistic path integral
for a
particle
in a potential $V({\bf x})$
\cite{7,1},
we find the expression for the fixed-energy amplitude
\begin{equation}
\label{3.1}G({\bf x}_b,{\bf x}_a;E)=\frac{i\hbar }{2Mc}\int_0^\infty dL\int
D\rho \Phi \left[ \rho \right] \int D^Dxe^{-A_E/\hbar }
\end{equation}
with the action
\begin{equation}
\label{3.2}A_E=\int_{\tau _a}^{\tau _b}d\tau \left[ \frac M{2\rho
\left( \tau \right) }{\bf x}^{\prime ^2}\left( \tau \right) -i\frac{e}{c}%
{\bf A({\bf x})\cdot x^{\prime }(}\tau {\bf )}-\rho (\tau )\frac{\left(
E-V\right) ^2}{2Mc^2}+\rho \left( \tau \right) \frac{Mc^2}2\right] .
\end{equation}
For the ABC system under consideration, the potential is
\begin{equation}
\label{3.6}V(r)=-e^2/r,
\end{equation}
and the vector potential
\begin{equation}
\label{3.7}A_i=2g\partial _i\theta ,
\end{equation}
where $e$ is the charge and $\theta $ is the azimuthal angle around the
tube:
\begin{equation}
\label{3.8}\theta ({\bf x})=\arctan (x_2/x_1).
\end{equation}
The associated magnetic field lines are confined to an infinitely thin tube
along the $z$-axis:
\begin{equation}
B_3=2g\epsilon _{3jk}\partial _j\partial _k\theta
=2g2\pi \delta ^{(2)}({\bf x}_{\bot }),
\label{3.9}\end{equation}
where ${\bf x}_{\bot }$ is the transverse vector ${\bf x}_{\bot }\equiv
(x_1,x_2).$

Before time-slicing the path integral, we have to regularize it
via a so-called $f$-transformation \cite{1,5}, which exchanges the path
parameter $ \tau$ by a new one $s$:
\begin{equation}
\label{3.3}d\tau =dsf_l({\bf {x}}_n)f_r({\bf {x}}_{n-1}),
\end{equation}
where $f_l({\bf {x}})$ and $f_r({\bf {x}})$ are
invertible functions whose product is positive. The freedom in choosing $%
f_{l,r}$ amounts to an invariance under path-dependent-reparametrizations
of the path parameter $ \tau$
in the fixed-energy amplitude  (\ref{3.1}). By
this transformation, the (D+1)-dimensional relativistic fixed-energy
amplitude for arbitrary time-independent potential turns into \cite{1,5}%
\begin{eqnarray}
&&\!\!\!\!\!\!\!\!\!\!\!\!\!\!\!\!\!\!\!\!G({\bf x}_b,{\bf x}_a;E)\approx
\frac{i\hbar }{2Mc}\int_0^\infty dS
 \prod_{n=1}^{N+1}\left[ \int d\rho _n\Phi (\rho _n)\right]
\nonumber \\
&&~~~~~~~~\times
\frac{f_l(%
{\bf {x}}_a)f_r({\bf {x}}_b)}{\left[ \lfrac{2\pi \hbar \epsilon _b^s\rho
_bf_l({\bf {x}}_b)f_r({\bf {x}}_a)}M\right] ^{D/2}} 
\prod_{n=1}^N\left[ \int_{-\infty }^\infty \frac{d^Dx_n}{\left[ \lfrac{2\pi
\hbar \epsilon _n^s\rho _nf({\bf {x}}_n)}M\right] ^{D/2}}\right] \exp
\left\{ - A^N/\hbar \right\}
\label{3.4}
\end{eqnarray}
with the $s$-sliced action%
$$ \!\!\!\!\!\!\!\!\!\!\!\!\!\!\!\!\!\!\!\!\!\!\!\!\!\!\!
A^N=\sum_{n=1}^{N+1}\left[ \frac{M\left( {\bf {x}}_n-{\bf {x}}_{n-1}\right)
^2}{2\epsilon _n^s\rho _nf_l({\bf {x}}_n)f_r({\bf {x}}_{n-1})}-i\frac{e}{c}{\bf
A}%
_n\cdot ({\bf x}_n-{\bf x}_{n-1})\right.
$$
\begin{equation}
\label{3.5}~~~~~~~~~~~~~~~~~~~\left. -\epsilon _n^s\rho _nf_l({\bf
{x}}_n)f_r({\bf {x}}_{n-1})
\frac{\left( E-V\right) ^2}{2Mc^2}+\epsilon _n^s\rho _nf_l({\bf {x}}_n)f_r({\bf
{x}}_{n-1})\frac{Mc^2}2\right] .
\end{equation}
A family
functions which regulates the ABC system is
\begin{equation}
\label{3.10}f_l({\bf {x}})=f({\bf x})^{1-\lambda },\ \ f_r({\bf {x}})=f({\bf %
x})^\lambda ,
\end{equation}
whose product satisfies $f_l({\bf {x}})f_r({\bf {x}})=f({\bf {x}})=r.$ Thus
arrive at
the amplitude
\begin{equation}
\label{3.11}
G({\bf x}_b,{\bf x}_a;E)\approx \frac{i\hbar }{2Mc}\int_0^\infty dS
\prod_{n=1}^{N+1}\left[ \int d\rho _n\Phi (\rho
_n)\right] 
\frac{%
\left( r_a/r_b\right) ^{1-2\lambda }}{{2\pi \hbar \epsilon
_b^s\rho _b}/M}
\prod_{n=2}^{N+1}\left[ \int_{-\infty }^\infty \frac{d^2\triangle
x_n}{ {2\pi \hbar \epsilon _n^s\rho _nr_{n-1}}/M }\right]
\exp \left\{ -\frac 1\hbar A^N\right\} ,
\end{equation}
with the action
\begin{equation}
\label{3.12}\left.
A^N=\sum_{n=1}^{N+1}\left[ \frac{M\left( {\bf {x}}_n-{\bf {x}}_{n-1}\right)
^2}{2\epsilon _n^s\rho _nr_n^{1-\lambda }r_{n-1}^\lambda }-i\frac {e}{c}{\bf A}_n\cdot (%
{\bf x}_n-{\bf x}_{n-1})\right.
-\epsilon _n^s\rho _nr_n\left( r_{n-1}/r_n\right)
^\lambda \frac{(E-V)^2}{2Mc^2}+\epsilon _n^s\rho _n\left( r_{n-1}/r_n\right)
^\lambda \frac{Mc^2}2\right] .
\end{equation}
Since the path integral represents the general relativistic resolvent
operator, all results must be independent of the splitting parameter $%
\lambda $ after going to the continuum limit. Choosing
$\lambda =1/2,$ we obtain the continuum limit
\begin{equation}
\label{3.13}A_E{\bf \left[ x,x^{\prime }\right] }=\int ds\left[ \frac{%
Mx^{\prime 2}}{2\rho r}-i\frac{e}{c}{\bf A}
\cdot {\bf x}^{\prime }-\rho r\frac{(E-V)^2}{2Mc^2}+\rho r
\frac{Mc^2}2\right] .
\end{equation}
We now solve the $s$-sliced ABC system
as in the case of the two-dimensionel Coulomb problem without the
Aharonov-Bohm potential \cite{1}.
We introducing the {\it Levi-Civit\'a}
transformation
\begin{equation}
\label{3.14}\left(
\begin{array}{r}
x_1 \\
x_2
\end{array}
\right) =\left(
\begin{array}{cc}
u^1 & -u^2 \\
u^2 & u^1
\end{array}
\right) \left(
\begin{array}{c}
u^1 \\
u^2
\end{array}
\right)
\end{equation}
and write this in a matrix form:
\begin{equation}
\label{3.15}{\bf x}=A({\bf u}){\bf u}.
\end{equation}
For every slice, the coordinate transformation reads
\begin{equation}
\label{3.16}{\bf x}_n=A({\bf u}_n){\bf u}_n
\end{equation}
yielding
\begin{equation}
\label{3.17}(\triangle {\bf x}_n^i)^2=4{\bf \bar u}_n^2(\triangle {\bf u}%
_n^i)^2,
\end{equation}
where ${\bf \bar u}_n$ $\equiv ({\bf u}_n+{\bf u}_{n-1})/2$.
For the sliced AB potential,
the {\it Levi-Civit\'a} transformation yields
\begin{equation}
\label{3.20}
{\bf A}_n\cdot ({\bf x}_n-{\bf x}_{n-1})=-2g\frac{(x_2)_n(\triangle
x_1)_n-(x_1)_n(\triangle x_2)_n}{r_n^2}
= -4g\frac{u_n^2\triangle u_n^1-u_n^1\triangle u_n^2}{%
{\bf u}_n^2} .
\end{equation}
Thus we obtain for the path integral (\ref{3.11})
the Duru-Kleinert-tranformed expression:
\begin{equation}
\label{3.21}G({\bf x}_b,{\bf x}_a;E)= \frac{i\hbar }{2Mc}\int_0^\infty
dS\ e^{SEe^2/\hbar Mc^2}\frac 14\left[ G({\bf u}_b,{\bf u}_a;S)+G(-{\bf u}_b,%
{\bf u}_a;S)\right] ,
\end{equation}
where $G({\bf u}_b,{\bf u}_a;S)$ is the $s$-sliced amplitude
of a harmonic oscillator
in an Aharonov-Bohm vector potential corresponding to twice the magnetic field
(\ref{3.9}) in ${\bf u}$-space:
$$
 G({\bf u}_b,{\bf u}_a;S)=\prod_{n=1}^{N+1}\left[ \int d\rho _n\Phi (\rho
_n)\right] \frac 1{
\lfrac{2\pi \hbar \epsilon _b^s\rho _b}M }\prod_{n=1}^N\left[
\int_{-\infty }^\infty \frac{d^2u_n}{ \lfrac{2\pi \hbar \epsilon
_n^s\rho _n}M }\right] \exp \left\{ -\frac 1\hbar A^N\right\}, 
$$
with the action
\begin{equation}
\label{3.22}A^N=\sum_{n=1}^{N+1}\left\{ \frac{m(\triangle {\bf u}_n)^2}{%
2\epsilon _n^s\rho _n}-2i\frac{e}{c}({\bf A}_n\cdot \triangle {\bf
u}_n)+\epsilon
_n^s\rho _n\frac{m\omega ^2{\bf u}_n^2}2-\epsilon _n^s\rho _n\frac{\hbar
^24\alpha ^2}{2m{\bf u}_n^2}\right\} .
\end{equation}
Here
\begin{equation}
m=4M,~~~~~~
\label{3.23}
\omega ^2=\frac{M^2c^4-E^2}{4M^2c^2},
\end{equation}
and
\begin{equation}
\label{3.23.1}{\bf A}_n\cdot \triangle {\bf u}_n=-2g\frac{u_n^2\triangle
u_n^1-u_n^1\triangle u_n^2}{{\bf u}_n^2}.
\end{equation}
The symmetrization in ${\bf u}_b$ in Eq. (\ref{3.21}) is necessary since for
each path from ${\bf x}_a$ to ${\bf x}_b$, there are two paths in
the square root space, one from ${\bf u}_a$ to ${\bf u}_b$ and one from $%
{\bf u}_a$ to $-{\bf u}_b$.

As in the two-dimensional Coulomb problem,
there are no $s$-slicing corrections \cite{1}.

Let us now
analyze the effect come form the magnetic interaction upon the Coulomb
system, defining the azimuthal angle
$
\varphi ({\bf u})=\arctan (u^2/u^1)=\theta({\bf x})/2$
in the $u$-plane, so that
$A_\mu =2g\partial _\mu \varphi,~
B_3=2g\epsilon _{3jk}\partial _j\partial _k\varphi$.
Note that
derivatives in front of $\varphi ({\bf u})$ commute everywhere,
except at the origin where Stokes' theorem yields
\begin{equation}
\label{3.25}\int d^2u(\partial _1\partial _2-\partial _2\partial _1)\varphi
=\oint d\varphi =2\pi
\end{equation}
The magnetic flux through the tube is defined by the integral
\begin{equation}
\label{3.26}\Phi =\int d^2uB_3.
\end{equation}
A comparison with the equation for $\varphi ({\bf u})$
shows that the coupling constant $g$
is related to the magnetic flux by
\begin{equation}
\label{3.27}g=\frac \Phi {4\pi }.
\end{equation}
When inserting $A_\mu =2g\partial _\mu \varphi $ into Eq. (\ref{3.22}), the
interaction takes the form
\begin{equation}
\label{3.28}A_{\rm mag}=-2\hbar \betan _0\int_0^Sds\varphi ^{\prime }(s),
\end{equation}
where $\varphi(s)\equiv  \varphi({\bf u}(s) )$, and
$\betan _0$ is the dimensionless number
\begin{equation}
\label{3.29}\betan _0\equiv -\frac{2eg}{\hbar c}.
\end{equation}
The minus sign is a matter of convention. Since the particle orbits are
present at all times, their worldlines in spacetime can be considered as
being closed at infinity, and the integral
\begin{equation}
\label{3.30}n=\frac 1{2\pi }\int_0^Sds\varphi ^{\prime }
\end{equation}
is the topological invariant with integer values of the winding number $n$.
The magnetic interaction is therefore a purely topological one, its value
being
\begin{equation}
\label{3.31}A_{\rm mag}=-\hbar \betan _04\pi n.
\end{equation}
After adding this to the action of Eq. (\ref{3.22}) in the radial
decomposition of the relativistic path integral \cite{5,6}, we rewrite the
sum over the azimuthal quantum numbers $k$ via Poisson's summation formula,
and obtain%
\begin{equation}
\label{3.32}
G({\bf u}_b,{\bf u}_a;S)=\int_{-\infty }^\infty d\betan \frac 1{\sqrt{u_bu_a}%
}G(u_b,u_a;S)_\betan
\times \sum_{n=-\infty }^\infty \frac 1{2\pi }e^{i(\betan -2\betan
_0)(\varphi _b+2n\pi -\varphi _a)}.
\end{equation}
Since the winding number $n$ is often not easy to measure experimentally,
let us extract observable consequences which are independent of $n.$ The sum
over all $n$ forces $\betan $ to be equal to $2\betan _0$ modula an arbitrary
integer number. the result is
\begin{equation}
\label{3.33}G({\bf u}_b,{\bf u}_a;S)=\sum_{k=-\infty }^\infty \frac 1{\sqrt{%
u_bu_a}}G(u_b,u_a;S)_{k+2\betan _0}\frac 1{2\pi }e^{ik(\varphi _b-\varphi
_a)}.
\end{equation}

We now choose the gauge $\rho (s)=1$ in Eq. (\ref{3.22}). This leads to the
Duru-Kleinert transformed action
\begin{equation}
\label{3.34}A^N=\int_0^Sds\left[ \frac{m{\bf u}^{\prime 2}}2-2i\frac{e}{c}({\bf
A}%
\cdot {\bf u^{\prime }})+\frac{m\omega ^2{\bf u}^2}2-\frac{4\hbar ^2\alpha
^2 }{2m{\bf u}^2}\right] .
\end{equation}
where $\alpha $ denotes the fine-structure constant $%
\alpha \equiv e^2/\hbar c\approx 1/137$.
This action describes a particle
of mass $%
m=4M$ moving as a function of the ``pseudotime'' $s$ in an
Aharonov-Bohm field and a harmonic
oscillator potential of frequency
\begin{equation}
\label{3.35}\omega ^2=\frac{Mc^2-E^2}{4M^2c^2}.
\end{equation}
In addition, there is
an extra attractive potential $V_{{\rm extra}}=-4\hbar
^2\alpha ^2/2m{\bf u}^2$ looking like an inverted
centrifugal barrier
which is comveniently parametrized with the help of a corresonding
angular momentum $l_{{\rm extra}}$,
whose square is
negative: $l_{{\rm extra}}^2\equiv -4\alpha ^2,$ writing
$V_{{\rm extra}}=\hbar ^2\lfrac{l_{{\rm extra}}^2}{2m{\bf u}^2}$.
Such an extra potential can easily be incorporated into
the amplitude of the pure Coulomb system
by a technique developed in the treatment of the
radial part of the harmonic oscillator path integral \cite{rad},
yielding a radial amplitude for the azimuthal quantum number $k$:
\begin{equation}
\label{3.37}G( u_b,u_a;S)_k=\frac m\hbar \frac{\omega
\sqrt{u_bu_a}}{\sinh
\omega s}e^{-\frac{m\omega }{2\hbar }\left( u_b^2+u_a^2\right) \coth \omega
s}I_{\sqrt{\mid k\mid ^2-4\alpha ^2}}\left( \frac m\hbar \frac{\omega u_bu_a
}{\sinh \omega s}\right) .
\end{equation}
where $I_\nu $ is the modified Bessel function. Incorporating also
the
effect of the Aharonov-Bohm potential yields
\begin{equation}
\label{3.38}G( u_b,u_a;S)_{k+2\betan _0}=\frac m\hbar \frac{\omega
\sqrt{u_bu_a%
}}{\sinh \omega s}e^{-\frac{m\omega }{2\hbar }\left( u_b^2+u_a^2\right)
\coth \omega s}I_{\sqrt{\mid k+2\betan _0\mid ^2-4\alpha ^2}}\left( \frac
m\hbar \frac{\omega u_bu_a}{\sinh \omega s}\right) .
\end{equation}
These radial amplitudes can now be combined with angular wave functions to
find the full amplitude (\ref{3.32}).

Inserting the result into the integral representation
(\ref{3.21})
for the resolvent,
we use
polar coordinates in ${\bf x}$-space
with $\theta =2\varphi ,r=u^2,$ and
obtain the
expression
\begin{equation}
\label{3.39}G({\bf x}_b,{\bf x}_a;E)=\sum_{k=-\infty }^\infty
G(r_b,r_a;E)_k \frac1{2\pi}e^{ik(\theta _b-\theta _a)},
\end{equation}
where%
\begin{equation}
\label{3.40}
G(r_b,r_a;E)_k=\frac {i\hbar} {2Mc}\frac{2M}\hbar \int_0^\infty
dS\;e^{e^2ES/\hbar Mc^2}
 \frac{\omega }{\sinh \omega s}e^{-\frac{%
m\omega }{2\hbar }\left( r_b+r_a\right) \coth \omega s}I_{\sqrt{4\mid
k+\betan _0\mid ^2-4\alpha ^2}}\left( \frac m\hbar \frac{\omega \sqrt
{r_br_a}}{%
\sinh \omega s}\right) .
\end{equation}
The integral can be calculated with the help of the formula%
\begin{equation}
\label{3.41}
\int_0^\infty dy\frac{e^{2\nu y}}{\sinh y}\exp \left[ -\frac t2\left( \zeta
_a+\zeta _b\right) \coth y\right] I_\mu \left( \frac{t\sqrt{\zeta _b\zeta _a}
}{\sinh y}\right)
=\frac{\Gamma \left( \left( 1+\mu \right) /2-\nu \right) }{t
\sqrt{\zeta _b\zeta _a}\Gamma \left( \mu +1\right) }W_{\nu ,\mu /2}\left(
t\zeta _b\right) M_{\nu ,\mu /2}\left( t\zeta _b\right) ,
\end{equation}
with the range of validity%
$$
\begin{array}{l}
\zeta _b>\zeta _a>0, \\
{Re}[(1+\mu )/2-\nu ]>0, \\ {Re}(t)>0,\mid \arg t\mid <\pi ,
\end{array}
$$
where $M_{\mu ,\nu }$ and $W_{\mu ,\nu }$ are the Whittaker functions.
In this way, we obtain the final result
for the radial amplitude
valid for $u_{b}>u_a$:
$$
G(r_b,r_a;E)_k=\frac {i\hbar }{2Mc}\frac{4Mc}{\sqrt{M^2c^4-E^2}}
\times \frac{\Gamma \left( 1/2+\sqrt{\mid k+\betan _0\mid ^2-\alpha ^2}%
-E\alpha /\sqrt{M^2c^4-E^2}\right) }{\sqrt{r_ar_b}\Gamma \left( 2\sqrt{\mid
k+\betan _0\mid ^2-\alpha ^2}+1\right) }
{}~~~~~~~~~~~~~~~~~~~~~~~~~~~~~~~~~~~~~~$$
\begin{equation}
\label{3.42}
{}~~~~~~\times W_{E\alpha /\sqrt{M^2c^4-E^2},\sqrt{\mid k+\betan _0\mid
^2-\alpha ^2}%
}\left( \frac 2{\hbar c}\sqrt{M^2c^4-E^2}r_b\right)
 M_{E\alpha /\sqrt{M^2c^4-E^2},\sqrt{\mid k+\betan _0\mid
^2-\alpha ^2}}\left( \frac 2{\hbar c}\sqrt{M^2c^4-E^2}r_a\right) .
\end{equation}
The energy spectra can be extracted from the poles. They are determined by
\begin{equation}
\label{3.43}\frac12+\sqrt{\mid k+\betan _0\mid ^2-\alpha ^2}-
\frac {E\alpha}{\sqrt{
M^2c^4-E^2}}=-n_r,~~~~~n_r=0,1,2,\cdots .
\end{equation}
Expanding this equation into powers of $\alpha ,$ we get
\begin{eqnarray}
\label{3.44}\left.
E_{nk}=\pm Mc^2 \left\{ \vbox to 24pt{}
 1\!-\!\frac 12\left[ \frac \alpha {n+\mid k+\betan _0\mid
\!-\!1/2}\right] ^2
\!-\!\frac{\alpha ^4}{\left( n+\mid k+\betan _0\mid \!-\!1/2\right)^3}%
\right.
 \left[ \frac 1{2\mid k+\betan _0\mid }\!-\!\frac
3{8\left( n+\mid k+\betan _0\mid \!-\!1/2\right) }\right] +\dots
%O\left( \alpha^6\right)
\vbox to 24pt{} \right\} ,
\end{eqnarray}
for $n=1,2,3,\cdots$.
In the non-relativistic limit,
the spectra reduces to that in Ref.~\cite{16,17,18}.

The alert reader will have noted the similarity of the
techniques used in this paper
to those leading to
the solution of the path integral of the dionium atom \cite{rad}.
\\
\\
\centerline{ACKNOWLEDGMENTS}
\\
\\
\centerline{The author is grateful to Professor  
H. Kleinert who 
critically read the entire manuscript and made corrections.}

\end {document}